\documentclass[preprint2,10.5pt]{aastex}
\begin{document}

\title{\large{\rm{NEW EVIDENCE SUPPORTING CLUSTER MEMBERSHIP FOR THE KEYSTONE CALIBRATOR $\delta$ CEPHEI}}}
\author{D. Majaess$^1$, D. Turner$^1$, W. Gieren$^2$}
\affil{$^1$ Saint Mary's University, Halifax, Nova Scotia, Canada}
\affil{$^2$ Universidad de Concepci\'on, Concepci\'on, Chile.}
\email{dmajaess@cygnus.smu.ca}

\begin{abstract}
New and existing $UBVJHK_s$, spectroscopic, NOMAD, HST, and revised HIP observations are employed to determine properties for $\delta$ Cep and its host star cluster.  The multi-faceted approach ensured that uncertainties were mitigated ($\sigma/d \sim 2$\%). The following fundamental parameters were inferred for $\delta$ Cep: $E(B-V)=0.073\pm0.018(\sigma )$, $\log{\tau}=7.9\pm0.1$, and $d=272\pm 3(\sigma_{\bar{x}}) \pm 5 (\sigma )$  pc.  The cluster exhibits a turnoff near B6 ($M_{*}/M_{\sun}\sim5$), and the brightest host cluster members are the supergiants $\zeta$ Cep (K1.5Ib) and $\delta$ Cep.  To within the uncertainties, the two stars share common astrometric parameters ($\pi$, $\mu_{\alpha}$, $\mu_{\delta}$, $RV\sim-17$ km/s) and are tied to bluer members via the evolutionary track implied by the cluster's $UBVJHK_s$ color-color and color-magnitude diagrams. The cluster's existence is bolstered by the absence of an early-type sequence in color-magnitude diagrams for comparison fields.   NOMAD data provided a means to identify potential cluster members ($n\sim30$) and double the existing sample.  That number could increase with forthcoming precise proper motions (DASCH) for fainter main-sequence stars associated with classical Cepheids (e.g., $\delta$ Cep), which may invariably foster efforts to strengthen the Galactic Cepheid calibration and reduce uncertainties tied to $H_0$.
\end{abstract}

\section{{\rm \footnotesize INTRODUCTION}}
Cepheid variables are utilized to establish extragalactic distances and constrain cosmological models \citep{mr09,ss10,ri11}.  However, the reliability of the derived parameters is invariably tied to the Cepheid calibration. \citet{fr01} noted that ambiguities related to the zero-point of the calibration account for a sizable fraction of the total uncertainty associated with $H_0$.  The uncertainty hinders efforts to constrain dark energy, since the parameter is acutely dependent on an accurate Hubble constant \citep[$\sigma_{\rm w}\sim2 \sigma_{\rm H_0}$,][]{mr09}.  The next generation follow-up to the HST key project to measure $H_0$ \citep[the Hubble Carnegie project,][]{fm10} aims to mitigate that problem by relying on LMC and Galactic calibrators \citep{be02,be07,tu10,st11}.  Consequently, bolstering the Galactic calibration should support efforts by the Hubble Carnegie, Araucaria, and S$H_0$ES projects to determine $H_0$ to within $2-4\%$ \citep[][]{gi05,ri11}.  

In this study, $UBVJHK_s$, spectroscopic, NOMAD, HST, and revised HIP observations for stars physically associated with $\delta$ Cep are employed to constrain its fundamental parameters: age ($\log{\tau}$), color-excess ($E_{B-V}$), distance, progenitor mass, and absolute Wesenheit magnitude ($W_{VI_c,0}$). 

\begin{deluxetable}{cccccccc}
\tablewidth{0pt}
\tabletypesize{\scriptsize}
\tablecaption{Cep OB6 Member List (Z99)}
\tablehead{\colhead{HIP ID} & \colhead{HIP $\pi$(mas)\tablenotemark{5}} & \colhead{V07 $\pi$(mas)\tablenotemark{5}} & \colhead{$\mu_{\alpha},\mu_{\delta}$ (mas/yr)\tablenotemark{6}} & \colhead{m.p.~(Z99)\tablenotemark{1}} & \colhead{m.p.\tablenotemark{2}} & \colhead{$E(B-V)$\tablenotemark{3}}}
\startdata
109426	&	$	3.8	\pm	0.7	$	&	$	3.6	\pm	0.5	$	& $	16.6	\pm	0.6	$ , $	4.1	\pm	0.5	$ &	94	&	m	&	--		\\
109492	&	$	4.5	\pm	0.5	$	&	$	3.9	\pm	0.1	$	& $	13.3	\pm	0.4	$ , $	4.4	\pm	0.3	$ &	97	&	m	&	--		\\
110266	&	$	3.9	\pm	0.6	$	&	$	4.4	\pm	0.3	$	& $	19.0	\pm	0.5	$ , $	5.1	\pm	0.5	$ &	96	&	m	&	 0.070 	\\
110275	&	$	4.0	\pm	1.0	$	&	$	4.0	\pm	0.9	$	& $	14.7	\pm	1.0	$ , $	5.7	\pm	0.9	$ &	89	&	m	&	--		\\
110356	&	$	3.4	\pm	0.7	$	&	$	2.9	\pm	0.5	$	& $	11.7	\pm	0.6	$ , $	3.2	\pm	0.6	$ &	100	&	m	&	0.085	\\
110497	&	$	3.8	\pm	0.6	$	&	$	3.2	\pm	0.4	$	& $	17.4	\pm	0.5	$ , $	4.8	\pm	0.5	$ &	98	&	m	&	0.060 	\\
110648	&	$	3.9	\pm	1.0	$	&	$	3.3	\pm	0.9	$	& $	16.2	\pm	1.0	$ , $	6.2	\pm	0.8	$ &	84	&	m	&		--	 \\
110807	&	$	4.0	\pm	0.6	$	&	$	3.5	\pm	0.4	$	& $	16.1	\pm	0.5	$ , $	5.4	\pm	0.4	$ &	92	&	m	&	0.060  	\\
110925	&	$	4.3	\pm	0.9	$	&	$	5.1	\pm	0.8	$	& $	21.3	\pm	1.1	$ , $	4.8	\pm	0.8	$ &	86	&	m	&	0.060\tablenotemark{4}	\\
110988	&	$	3.4	\pm	0.6	$	&	$	3.7	\pm	0.5	$	& $	16.4	\pm	0.7	$ , $	4.7	\pm	0.7	$ &	100	&	m	&	0.085\tablenotemark{4}	\\
111060	&	$	5.0	\pm	0.8	$	&	$	5.3	\pm	0.7	$	& $	17.2	\pm	0.7	$ , $	4.5	\pm	0.7	$ &	100	&	m	&	--		\\
112141	&	$	3.2	\pm	0.8	$	&	$	3.4	\pm	0.6	$	& $	14.6	\pm	0.7	$ , $	2.1	\pm	0.6	$ &	89	&	m	&	--		\\
113255	&	$	4.3	\pm	0.7	$	&	$	4.5	\pm	0.5	$	& $	19.8	\pm	0.6	$ , $	3.0	\pm	0.6	$ &	99	&	m	&	0.095	\\
113316	&	$	3.2	\pm	0.7	$	&	$	3.6	\pm	0.4	$	& $	14.1	\pm	0.7	$ , $	3.2	\pm	0.6	$ &	99	&	m	&	 0.100	\\
$\delta$ Cep	&	$	3.3	\pm	0.6	$	&	$	3.8	\pm	0.2	$	& $	16.4	\pm	0.6	$ , $	3.5	\pm	0.6	$ &	89	&	m	&	--	\\
110459	&	$	4.1	\pm	0.9	$	&	$	4.5	\pm	0.7	$	& $	16.2	\pm	0.9	$ , $	5.2	\pm	0.7	$ &	100	&	nm	&	--	\\
111069	&	$	3.2	\pm	0.9	$	&	$	3.1	\pm	0.8	$	& $	15.1	\pm	0.8	$ , $	6.9	\pm	0.7	$ &	79	&	nm	&	--		\\
112473	&	$	3.6	\pm	0.8	$	&	$	5.0	\pm	0.7	$	& $	13.9	\pm	0.8	$ , $	2.8	\pm	0.8	$ &	97	&	nm	&	--		\\
112998	&	$	2.5	\pm	0.6	$	&	$	2.6	\pm	0.3	$	& $	12.7	\pm	0.6	$ , $	2.2	\pm	0.5	$ &	98	&	nm	&		--	\\
113993	&	$	3.8	\pm	0.7	$	&	$	3.7	\pm	0.4	$	& $	14.3	\pm	0.6	$ , $	3.9	\pm	0.5	$ &	79	&	nm	&	 --		\\

\enddata
\tablenotetext{1}{\scriptsize{Membership probability assigned by \citet[][Z99]{ze99}.}}
\tablenotetext{2}{\scriptsize{Membership inferred from $UBVJHK_s$ and spectroscopic observations (Figs.~\ref{fig-ubv},~\ref{fig-cmd}).}}
\tablenotetext{3}{{\scriptsize Reddenings derived from the $UBV$ color-color analysis (Fig.~\ref{fig-ubv}).}}
\tablenotetext{4}{{\scriptsize Stars in close proximity to $\delta$ Cep.}}
\tablenotetext{5}{{\scriptsize Parallaxes from \citet[][HIP]{pe97} and \citet[][V07]{vle07}.}}
\tablenotetext{6}{{\scriptsize Proper motion data from NOMAD \citep{za04}.}}
\label{table1}
\end{deluxetable}

\section{{\rm \footnotesize ANALYSIS}}
\subsection{{\rm \footnotesize REVISED HIP OBSERVATIONS FOR CEP OB6}} 
\label{s-rhd}
\begin{figure*}[!t]
\epsscale{1.1}
\plotone{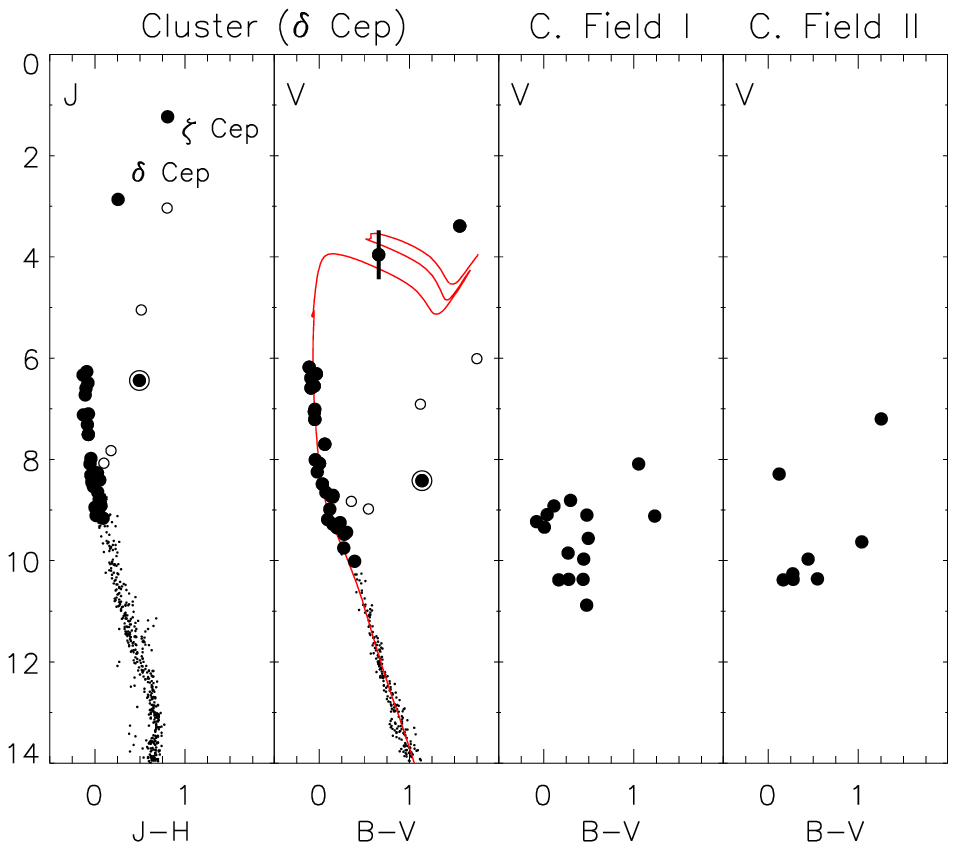}
\caption{\small{Left to right, $JH$ and $BV$ color-magnitude diagrams for the Table~\ref{table1} and NOMAD samples (panels 1, 2), and comparison fields (panels 3, 4).   \emph{Small dots} denote calibration stars from \citet{ma11}, which were employed to tie the cluster distance to a geometrically anchored scale \citep{vl09,ma11}. Large dots characterize stars exhibiting $\mu_{\alpha}=11-19$ and $\mu_{\delta}=2-7$ mas/yr.  Open circles are likely field stars (Table~\ref{table1}).  HIP 110459 (circled dot) was previously identified as a cluster member, yet $BVJHK_s$ photometry implies the object is a field star.  Panel 2, a Padova $\log{\tau}=7.9$ isochrone was applied.  The brightest cluster members are likely the supergiants $\zeta $ Cep (K1.5Ib) and $\delta$ Cep (amplitude variation indicated).  An early type cluster sequence is absent from the comparison fields (HIP data for the cluster $\mu_{\alpha}$/$\mu_{\delta}$).}}
\label{fig-cmd}
\end{figure*}

In their comprehensive study, \citet{ze99} discovered that $\delta$ Cep was a member of a group denoted as Cep OB6 \citep[see also][]{ho97}. 20 stars identified by \citet{ze99} as Cep OB6 members are highlighted in Table~\ref{table1}.  Stars exhibiting spectral types inconsistent with cluster membership based on their $UBVJHK_s$ color-color and color-magnitude positions were flagged as probable non-members. For example, HIP 110459 was previously assigned a membership probability of 100\% (Table~\ref{table1}), yet the star exhibits $JHK_s$ photometry and a late-type temperature class \citep[K5,][]{sk10} indicative of a field red clump giant (Fig.~\ref{fig-cmd}). $UBVJHK_s$ photometry was obtained from \citet{me91} and 2MASS \citep{cu03}.  Spectral types were assigned to stars featured in the Catalogue of Stellar Spectral Classifications \citep{sk10}. 

Revised HIP parallaxes \citep{vle07} were tabulated for 15 stars in Table~\ref{table1} which were identified as probable cluster members.  The revised HIP parallaxes exhibit a $\sim30$\% reduction in uncertainties relative to existing data \citep{pe97}, and the parallaxes for $\delta$ Cep and HD 213307 ($r \sim 0.7 \arcmin$) were increased from $\pi=3.32\pm0.58$:$3.43\pm0.64$ to $\pi=3.77\pm0.16$:$3.69\pm0.46$ mas.  A mean distance for the revised cluster sample outlined in Table~\ref{table1} is $d=271\pm11(\sigma_{\bar{x}})\pm42(\sigma)$ pc \citep[see also][and their appendix B]{ze99}.   

\subsection{{\rm \footnotesize REDDENING}} 
$UBVJHK_s$ color-color analyses permit an assessment of the sample's extinction properties.  $UBV$ data are particularly efficient at identifying early type stars owing to the $U$-band's sensitivity to the Balmer decrement \citep[e.g.,][]{tu89,ca06}.   

\begin{figure}[!t]
\epsscale{0.7}
\plotone{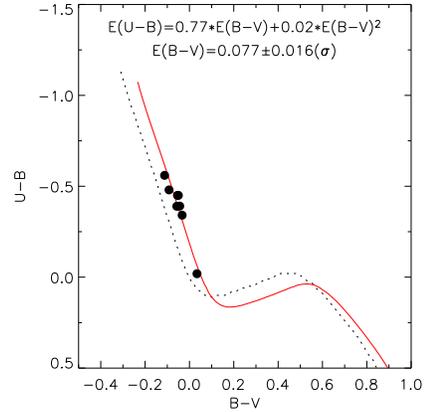}
\caption{\small{color-color diagram for stars in Table~\ref{table1} which are associated with $\delta$ Cep and possess $UBV$ photometry.  The sample is offset from the intrinsic relation (dotted line) by $E(B-V)=0.077\pm0.016(\sigma)$ (solid line).  The result confirms the reddening established for $\delta$ Cep by \citet{be02}.  The intrinsic relation and reddening law for the region were adopted from \citet{tu76,tu89}.}}
\label{fig-ubv}
\end{figure}

Cluster members featuring $UBV$ photometry \citep{me91} are offset from the intrinsic $UBV$ color-color relation by $E(B-V)=0.077\pm0.016 (\sigma)$ (Table~\ref{table1}, Fig.~\ref{fig-ubv}).  The finding supports the field reddening determined for $\delta$ Cep by \citet{be02}, and are consistent with those established from spectroscopic and $JHK_s$ observations.  However, uncertainties associated with the latter hamper a precise assessment.  The minimal spread, $E(B-V)\sim0.06-0.10$ (Table~\ref{table1}), may be indicative of marginal differential reddening, rotation, binarity, or photometric uncertainties.

\subsection{{\rm \footnotesize AGE}} 
$UBVJHK_s$ color-color analyses imply a cluster turnoff near B5-B7 ($M_{*}/M_{\sun}\sim5$), according to intrinsic relations \citep[Padova models, and][]{sl09,tu76,tu89,tu11}.  A $\log{\tau}=7.9\pm0.1$ isochrone was subsequently adopted based on the turnoff determined (Fig.~\ref{fig-ubv}, Table~\ref{table1}), and since the implied evolutionary track (Fig.~\ref{fig-cmd}) matches bluer and redder evolved cluster members ($\delta$ and $\zeta$ Cep). The result agrees with the Cepheid's predicted age \citep{tu96,bo05}.  The temporal match is pertinent evidence in support of cluster membership for $\delta$ Cep. 

\subsection{{\rm \footnotesize CLUSTER DISTANCE}} 
A precise distance may be established since two of four principal parameters associated with isochrone fitting were constrained by the $UBVJHK_s$ color-color and spectroscopic analyses, namely the reddening and age (spectral type at the turnoff).  $\delta$ Cep exhibits solar abundances, and hence the remaining parameter is the shift required in magnitude space to overlay the intrinsic relation upon the data. The resulting distance is $d=277\pm15$ pc (Fig.~\ref{fig-cmd}).   The zero-point is tied to seven benchmark open clusters ($d<250$ pc) which exhibit matching $JHK_s$ and revised Hipparcos distances \citep[the Hyades, $\alpha$ Per, Praesepe, Coma Ber, IC 2391, IC 2609, and NGC 2451,][]{vl09,ma11}.  A redetermination of the HST parallax for the Hyades supports that scale \citep{mc11}.  Isochrones, models, and the distance scale should be anchored (\& evaluated) using clusters where consensus exists, rather than the discrepant case (i.e. the Pleiades).  A ratio of total to selective extinction $R_J$ was adopted from \citet{ma11b} \citep[see also][]{bo04}, whereas a value for $R_V$ was adopted from \citet{tu76}. An advantage of employing  $JHK_s$ observations is that the cluster reddening is negligible in that part of the spectrum \citep[$E_{J-H}\sim0.3 \times E_{B-V}$,][and references therein]{ma08,bo04}, which consequently mitigates the impact of uncertainties in $R_{\lambda}$ ($J_0=J-E_{J-H} \times R_J$).

The distance derived from the cluster color-magnitude diagrams (Fig.~\ref{fig-cmd}) is tied to additional potential cluster members identified using  
NOMAD \citep{za04}.  That database was queried for stars exhibiting $\mu_{\alpha}=11-19$ and $\mu_{\delta}=2-7$ mas yr$^{-1}$ (Table~\ref{table1}).  Stars fainter than $J\sim9.2$ were culled from the resulting sample to reduce field contamination.  Proper motions may be less reliable for such stars, and spectroscopic and $UBV$ observations are typically unavailable.  Stars redder than $J-H\sim0.4$ were likewise removed to mitigate contamination from field red clump giants.  In addition, stars featuring anomalous positions in the multiband color-color and color-magnitude diagrams were removed.  The remaining stars double the number of existing potential cluster members, and are highlighted in Table~\ref{table2}.  The stars in Tables~\ref{table1} and \ref{table2} are potential members pending further evidence.  The NOMAD proper motions are consistent with estimates from the PPMXL catalog (Table~\ref{table2}). $UBVJHK_s$ photometry was sought from \citet{me91}, 2MASS \citep{cu03}, and pertinent resources.  For example, new observations acquired from the Bright Star Monitor (BSM), which is part of the AAVSO's robotic telescope network, provided $BV$ photometry for HD 239949: $V=10.013\pm0.031$ and $B-V=0.392\pm0.048$.  The cluster reddening was redetermined ($E_{B-V}=0.073\pm0.018 (\sigma)$) using four of the earlier type stars highlighted in Table~\ref{table2} which possess $UBV$ photometry.  

\begin{figure}[!t]
\epsscale{0.7}
\plotone{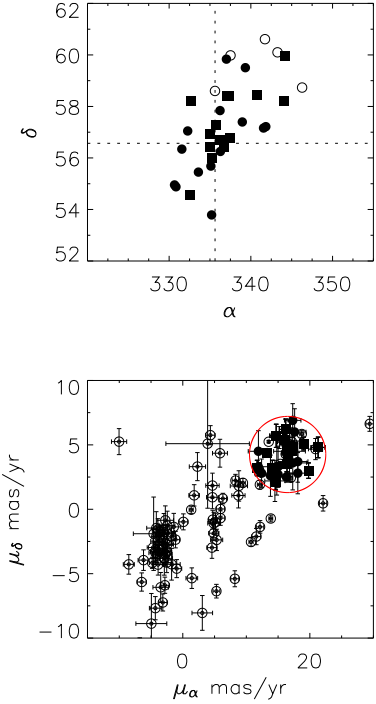}
\caption{\small{Top, J2000 RA/DEC positions for probable (squares) and non-members (open circles) highlighted in Table~\ref{table1}, and new potential members (filled circles) outlined in Table~\ref{table2}.  Bottom, dotted open circles represent \textit{all} HIP stars near the (approximate) cluster center.}}
\label{fig-pm}
\end{figure}

Cluster members appear to aggregate near J2000 coordinates of $22^{\rm h} 22.5^{\rm m} +56{\rm \degr} 34{\rm \arcmin}$ (Fig.~\ref{fig-pm}).  $\delta$ Cep lies at the periphery of the density enhancement, and within the confines of the cluster since the corona extends further \citep{kh69,tu85}.  The Cepheid is $r\sim 2 \degr$ from the aforementioned coordinates, which is equivalent to a linear projected separation of $\sim 9$ pc.  The revised HIP and HST parallaxes for $\delta$ Cep and HD 213307 ($r \sim 0.7 \arcmin$ from $\delta$ Cep), in tandem with their apparent positions, suggest that the distance to the cluster center and Cepheid is analogous to within the uncertainties.

\subsection{{\rm \footnotesize MEAN DISTANCE TO $\delta$ CEP}} 
The mean HIP parallax for the cluster (\S \ref{s-rhd}) agrees with the HIP parallax for $\delta$ Cep ($\pi=3.77\pm0.16$ mas), the HST parallax for $\delta$ Cep \citep[$\pi=3.66\pm0.15$ mas,][]{be02}, and the distance inferred for the host cluster from $UBVJHK_s$ and spectroscopic observations (Figs.~\ref{fig-ubv},~\ref{fig-cmd}).  Assigning equal weight to each method, the mean of the four distances for $\delta$ Cep is: $d=272\pm 3(\sigma_{\bar{x}}) \pm 5 (\sigma )$ pc.  That result agrees with the \citet{st11} determination from the infrared surface brightness technique \citep[IRSB,][]{fg97}.  The associated standard error and deviation provide a realistic estimate for the systematic uncertainty, which is often difficult to characterize.  

The resulting $VI_c$ Wesenheit magnitude for $\delta$ Cep is $W_{VI_c,0}=-5.12$ ($R_{VI_c}=2.55$).  That is consistent with results established for classical Cepheids displaying similar pulsation periods: CV Mon, V Cen, Y Sgr, and CS Vel \citep{be07,tu10,ma11b}.

\section{{\rm \footnotesize CONCLUSION \& FUTURE RESEARCH}}
The evidence presented bolsters the assertion by \citet{ze99} that $\delta$ Cep is a constituent of an intermediate age cluster.   The brightest cluster member is the K1.5Ib supergiant $\zeta$ Cep.  $\delta$ and $\zeta$ Cep share similar HIP parallaxes ($\pi=3.77\pm0.16$:$3.90\pm0.10$ mas), proper motions, radial velocities ($RV\sim-17$ km/s), and evolutionary histories (Fig.~\ref{fig-cmd}).   In tandem with the aforementioned evidence, the cluster's existence is supported by the absence of early-type stars from the comparison fields (Fig~\ref{fig-cmd}). NOMAD data were employed to identify additional potential cluster members (Tables~\ref{table1},~\ref{table2}). 

 The Cepheid exhibits parameters of: $E(B-V)=0.073\pm0.018 (\sigma)$, $\log{\tau}=7.9\pm0.1$, and $d=272\pm 3(\sigma_{\bar{x}}) \pm 5 (\sigma )$ pc (Table~\ref{table1}, Figs.~\ref{fig-ubv},~\ref{fig-cmd}).  The results are tied in part to spectroscopic and $UBVJHK_s$ observations, and may be adopted to refine classical Cepheid period-color, period-age, period-mass, period-luminosity, and period-Wesenheit relations \citep[e.g.,][]{tu10}.

Potential future research entails establishing precise proper motions for fainter stars near Cepheids using photographic plates stored at the CfA \citep[][DASCH]{gr07},\footnote{Digital Access to a Sky Century @ Harvard (DASCH), \url{http://hea-www.harvard.edu/DASCH/}} thereby extending the astrometric coverage provided by HIP.  The plate collection at the CfA offers unmatched multi-epoch observations spanning a $\sim100$ year temporal baseline. A concurrent venture pertains to employing new VVV $JHK_s$ observations for young clusters in Galactic spiral arms to calibrate adjacent long-period classical Cepheids \citep{mi10,mb11,ma11b}.  The aforementioned initiatives, in harmony with the analysis presented here, shall complement a suite of diverse efforts aimed at reducing uncertainties associated with $H_0$ in order to constrain cosmological models \citep[e.g.,][]{be07,fe08,ge11,ng11,sm11}.   

The agreement between the distances inferred for $\delta$ Cep from cluster membership and the IRSB technique suggests that the systematic uncertainties have been marginalized.  An analysis of $n=20$ cluster Cepheids featuring revised IRSB distances \citep{st11} yields a mean fractional difference of $-3\pm3$\%.  That result is reassuring, and subsequent research on the discrepant Cepheid calibrators (e.g., S Vul, SU Cas, in prep.) may reduce the remaining offset.  Further research is likewise required on the candidate cluster members associated with $\delta$ Cep (Tables~\ref{table1},~\ref{table2}).

\begin{deluxetable}{lcccccccc}
\tablewidth{0pt}
\tabletypesize{\scriptsize}
\tablecaption{Additional Potential Cluster Members}
\tablehead{\colhead{ID} & \colhead{$V$} & \colhead{$B-V$} & \colhead{$U-B$} & \colhead{$J$} & \colhead{$H$} & \colhead{$K_s$} & \colhead{$\mu_{\alpha},\mu_{\delta}$ (mas/yr)\tablenotemark{2}} & \colhead{$\mu_{\alpha},\mu_{\delta}$ (mas/yr)\tablenotemark{3}}}
\startdata
HD 210071\tablenotemark{1}            	&	6.39	&	-0.10	&	-0.45	&	6.49	&	6.57	&	6.58	&	$	16.3	\pm	0.4	$ , $	2.5	\pm	0.4	$	& $	16.3	\pm	0.4	$ , $	2.4	\pm	0.4	$	\\
HD 209636\tablenotemark{1} 	&	7.01	&	-0.05	&	-0.23	&	7.10	&	7.17	&	7.16	&	$	15.0	\pm	0.5	$ , $	2.7	\pm	0.5	$	& $	15.0	\pm	0.5	$ , $	2.6	\pm	0.5	$	\\
HD 214259 	&	8.72	&	0.15	&	0.10	&	8.27	&	8.24	&	8.24	&	$	18.1	\pm	1.6	$ , $	3.7	\pm	1.6	$	& $	18.4	\pm	1.3	$ , $	4.3	\pm	1.3	$	\\
HD 214512	&	8.74	&	0.15	&	--	&	8.30	&	8.29	&	8.29	&	$	17.6	\pm	1.8	$ , $	6.0	\pm	1.8	$	& $	-21.5	\pm	1.6	$ , $	53.1	\pm	1.6	$	\\
HD 212093            	&	8.25	&	-0.02	&	-0.29	&	8.31	&	8.36	&	8.35	&	$	15.0	\pm	1.6	$ , $	3.3	\pm	1.6	$	& $	14.1	\pm	1.2	$ , $	2.5	\pm	1.2	$	\\
HD 210480            	&	8.71	&	0.15	&	0.08	&	8.41	&	8.35	&	8.31	&	$	14.5	\pm	1.6	$ , $	2.2	\pm	1.6	$	& $	14.9	\pm	1.3	$ , $	2.3	\pm	1.3	$	\\
HD 211226            	&	8.65	&	0.07	&	0.03	&	8.45	&	8.49	&	8.47	&	$	17.3	\pm	1.3	$ , $	6.9	\pm	1.3	$	& $	15.4	\pm	1.2	$ , $	6.7	\pm	1.2	$	\\
HD 215879 	&	8.98	&	0.12	&	--	&	8.65	&	8.62	&	8.59	&	$	13.8	\pm	1.6	$ , $	2.6	\pm	1.6	$	& $	13.4	\pm	1.3	$ , $	4.4	\pm	1.2	$	\\
HD 212711 	&	9.25	&	0.23	&	--	&	8.78	&	8.72	&	8.66	&	$	17.3	\pm	2.2	$ , $	3.6	\pm	2.1	$	& $	16.7	\pm	1.9	$ , $	5.2	\pm	1.9	$	\\
HD 240052	&	9.44	&	0.30	&	--	&	8.79	&	8.74	&	8.71	&	$	18.1	\pm	1.6	$ , $	2.8	\pm	1.6	$	& $	17.3	\pm	1.3	$ , $	4.1	\pm	1.2	$	\\
HD 212137	&	9.19	&	0.09	&	--	&	8.95	&	8.95	&	8.94	&	$	12.4	\pm	1.1	$ , $	2.8	\pm	1.1	$	& $	12.9	\pm	1.2	$ , $	1.8	\pm	1.2	$	\\
BD+54$\degr$2675 	&	9.48	&	0.28	&	--	&	9.04	&	9.01	&	8.94	&	$	11.9	\pm	1.6	$ , $	4.5	\pm	1.6	$	& $	12.2	\pm	1.8	$ , $	7.5	\pm	1.7	$	\\
HD 239949                  	&	10.01	&	0.39	&	--	&	9.11	&	9.10	&	9.06	&	$	16.8	\pm	2.5	$ , $	4.5	\pm	2.3	$	& $	5.4	\pm	1.6	$ , $	-0.8	\pm	1.5	$	\\
BD+59$\degr$2523 	&	9.75	&	0.27	&	--	&	9.16	&	9.07	&	9.11	&	$	14.8	\pm	2.3	$ , $	3.3	\pm	2.2	$	& $	15.3	\pm	1.6	$ , $	4.2	\pm	1.6	$	\\
\enddata
\label{table2}
\tablenotetext{1}{\scriptsize{$\pi=5.06\pm0.33:5.54\pm0.39$ mas \citep{vle07}.}}
\tablenotetext{2}{\scriptsize{NOMAD proper motions \citep{za04}.}}
\tablenotetext{3}{\scriptsize{The PPMXL catalog \citep{ro10}.}}
\end{deluxetable}

\subsection*{{\rm \scriptsize ACKNOWLEDGEMENTS}}
\scriptsize{DM is grateful to the following individuals and consortia whose efforts lie at the foundation of the research: F. van Leeuwen \& M. Perryman (HIP), F. Benedict (HST), P. de Zeeuw, R. Hoogerwerf, J-C. Mermilliod, B. Skiff, NOMAD, 2MASS, AAVSO (T. Krajci, A. Henden, M. Simonsen), and staff at the CDS, arXiv, and NASA ADS.  WG gratefully acknowledges financial support for this work from the Chilean Center for Astrophysics FONDAP 15010003, and from the BASAL Centro de Astrofisica y Tecnologias Afines (CATA) PFB-06/2007. }


\begin{thebibliography}{}\setlength{\itemsep}{-1.5mm}
\bibitem[Benedict et al.(2002)]{be02} Benedict, G.~F., McArthur, B. E., Fredrick, L. W., et al.\ 2002, \aj, 124, 1695 
\bibitem[Benedict et al.(2007)]{be07} Benedict, G.~F., McArthur, B. E., Feast, M. W., et al., 2007, AJ, 133, 1810 
\bibitem[Bonatto et al.(2004)]{bo04} Bonatto, C., Bica, E., \& Girardi, L.\ 2004, \aap, 415, 571 
\bibitem[Bono et al.(2005)]{bo05} Bono, G., Marconi, M., 
Cassisi, S., et al.\ 2005, \apj, 621, 966 
\bibitem[Carraro et al.(2006)]{ca06} Carraro, G., Chaboyer, 
B., \& Perencevich, J.\ 2006, \mnras, 365, 867 
\bibitem[Cutri et al.(2003)]{cu03} Cutri, R.~M., Skrutskie, M. F., van Dyk, S., et al.\ 2003, The IRSA 2MASS All-Sky Point Source Catalog, NASA/IPAC Infrared Science Archive.
\bibitem[de Zeeuw et al.(1999)]{ze99} de Zeeuw, P.~T., 
Hoogerwerf, R., de Bruijne, J.~H.~J., Brown, A.~G.~A., 
\& Blaauw, A.\ 1999, \aj, 117, 354 
\bibitem[Feast(2008)]{fe08} Feast, M.~W.\ 2008, First Middle 
East-Africa, Regional IAU Meeting, held 5-10 April, 2008 in Cairo, 
Egypt (arXiv:0806.3019).
\bibitem[Fouque 
\& Gieren(1997)]{fg97} Fouque, P., \& Gieren, W.~P.\ 1997, \aap, 320, 799 
\bibitem[Freedman et al.(2001)]{fr01} Freedman, W.~L., 
Madore, B.~F., Gibson, B.~K., et al.\ 2001, \apj, 553, 47 
\bibitem[Freedman \& Madore(2010)]{fm10} Freedman, W.~L., \& Madore, B.~F.\ 2010, \araa, 48, 673 
\bibitem[Gerke et al.(2011)]{ge11} Gerke, J.~R., Kochanek, C.~S., Prieto, J.~L., Stanek, K.~Z., \& Macri, L.~M.\ 2011, arXiv:1103.0549 
\bibitem[Gieren et al.(2005)]{gi05} Gieren, W., 
Pietrzy{\'n}ski, G., Soszy{\'n}ski, I., et al.\ 2005, \apj, 628, 695 
\bibitem[Grindlay(2007)]{gr07} Grindlay, J.~E.\ 2007, The 
Central Engine of Active Galactic Nuclei, 373, 711 
\bibitem[Hoogerwerf et al.(1997)]{ho97} Hoogerwerf, R., de Bruijne, J.~H.~J., Brown, A.~G.~A., et al.\ 1997, Hipparcos - Venice '97, 402, 571 
\bibitem[Kholopov(1969)]{kh69} Kholopov, P.~N.\ 1969, 
\sovast, 12, 625 
\bibitem[Macri \& Riess(2009)]{mr09} Macri, L.~M., \& Riess, A.~G.\ 2009, American Institute of Physics Conference Series, 1170, 23 
\bibitem[Majaess et al.(2008)]{ma08} Majaess D.~J., Turner D.~G., Lane D.~J., 2008, MNRAS, 390, 1539
\bibitem[Majaess et al.(2011a)]{ma11} Majaess, D.~J., Turner, D.~G., Lane, D.~J., \& Krajci, T.\ 2011 (a), JAAVSO, in press (arXiv:1102.1705)
\bibitem[Majaess et al.(2011b)]{ma11b} Majaess, D., Turner, 
D., Moni Bidin, C., et al.\ 2011 (b), \apjl, 741, L27 
\bibitem[McArthur et al.(2011)]{mc11} McArthur, B.~E., 
Benedict, G.~F., Harrison, T.~E., \& van Altena, W.\ 2011, \aj, 141, 172 
\bibitem[Mermilliod(1991)]{me91} Mermilliod, J-C.\ 1991, Homogeneous Means in the UBV System, VizieR catalog.
\bibitem[Minniti et al.(2010)]{mi10} Minniti, D., Lucas, P. W., Emerson, J. P., et al.\ 2010, New Astronomy, 15, 433 
\bibitem[Moni Bidin et 
al.(2011)]{mb11} Moni Bidin, C., Mauro, F., Geisler, D., et al.\ 2011, \aap, 535, A33 
\bibitem[Ngeow(2011)]{ng11} Ngeow, C.-C.\ 2011, Proceedings of the 9th Pacific Rim Conference on Stellar Astrophysics (PRCSA2011), Lijiang, China, April 2011
(arXiv:1111.2094)
\bibitem[Perryman \& ESA(1997)]{pe97} Perryman, M.~A.~C., \& ESA 1997, ESA Special Publication, 1200
\bibitem[Riess et al.(2011)]{ri11} Riess, A.~G., Macri, L., 
Casertano, S., et al.\ 2011, \apj, 730, 119 
\bibitem[Roeser et al.(2010)]{ro10} Roeser, S., Demleitner, 
M., \& Schilbach, E.\ 2010, \aj, 139, 2440 
\bibitem[Shappee 
\& Stanek(2011)]{ss10} Shappee, B.~J., \& Stanek, K.~Z.\ 2011, \apj, 733, 124 
\bibitem[Skiff(2010)]{sk10} Skiff, B.\ 2010, General Catalogue of Stellar Spectral Classifications, VizieR catalog.
\bibitem[Steer \& Madore(2011)]{sm11} Steer, I., Madore, B., 2011, NED-D: a Master List of Extragalactic Distances, http://ned.ipac.caltech.edu/Library/Distances/ 
\bibitem[Storm et 
al.(2011)]{st11} Storm, J., Gieren, W., Fouqu{\'e}, P., et al.\ 2011, \aap, 534, A94 
\bibitem[Strai{\v z}ys \& Lazauskait{\.e}(2009)]{sl09} Strai{\v z}ys, V., \& Lazauskait{\.e}, R.\ 2009, Baltic Astronomy, 18, 19 
\bibitem[Turner(1976)]{tu76} Turner, D.~G.\ 1976, \aj, 81, 1125 
\bibitem[Turner(1985)]{tu85} Turner, D.\ 1985, IAU 
Colloq.~82: Cepheids: Theory and Observation, 209 

\bibitem[Turner(1989)]{tu89} Turner, D.~G.\ 1989, \aj, 98, 
2300 
\bibitem[Turner(1996)]{tu96} Turner, D.~G.\ 1996, \jrasc, 90, 82 
\bibitem[Turner(2010)]{tu10} Turner, D.~G.\ 2010, \apss, 326, 219 
\bibitem[Turner(2011)]{tu11} Turner, D.~G.\ 2011, RMxAA, 47, 127
\bibitem[van Leeuwen(2007)]{vle07} van Leeuwen, F.\ 2007, \aap, 474, 653 
\bibitem[van Leeuwen(2009)]{vl09} van Leeuwen, F.\ 2009, \aap, 497, 209 
\bibitem[Zacharias et al.(2004)]{za04} Zacharias, N., Monet, 
D.~G., Levine, S.~E., et al.\ 2004, Bulletin of the American Astronomical 
Society, 36, 1418 
\end{thebibliography}
\end{document}